
\input harvmac

\Title{LA-UR-92-2753}
{\vbox{\centerline{DYNAMICAL APPROACH TO PAIR}
\vskip2pt\centerline{PRODUCTION FROM STRONG FIELDS}}}
\vskip
.2in
\centerline{Fred Cooper}
\bigskip
\centerline{Theoretical Division T-8}
\bigskip
\centerline{Los Alamos National Laboratory}
\bigskip
\centerline{MS B-285 Los Alamos, NM 87545 USA}
\bigskip
\centerline{Lectures given at the NATO ADVANCED STUDY INSTITUTE}
\centerline{Particle Production in Highly Excited Matter}
\centerline{II Ciocco, Italy July 1992}
\Date{8/27/92}
\newsec{INTRODUCTION}

In  relativistic heavy-ion collisions one is hoping to produce conditions
where energy densities are high enough so that a new state of
matter-- the quark-gluon plasma can be produced. This state
of matter lasts for a short period of time following the collision
and may or may not be in equilibrium. Following this phase a
transition to  ordinary hadronic matter takes place  and many of the
processes which occur during the quark-gluon plasma phase might
be masked by processes which occur in the hadronic phase. In order
to determine processes which might be signals of the quark gluon
plasma one needs to know the dynamical evolution of the plasma.
This is because the  particles  that get produced during that phase have to
travel through a time evolving plasma.  In order to study this
problem one needs a different way of thinking about field theory.
Traditionally experiments in Elementary Particle Physics are black box
experiments where initial particles enter a region, final particles exit the
experimental region and all that is asked is how many
particles of what type, energy, etc. enter various detectors.  This type of
experiment  requires only a covariant S-Matrix
theory to predict the probabilities to be expected in the detectors. However,
if we want to know signatures of the quark
gluon plasma,we actually need to follow the time evolution of the plasma
and fields produced following  the heavy ion collision. This requires a
non-covariant real time formalism for the time evolution of  the quantum
fields.  In these talks we would first like to discuss various formalisms
for  doing real time calculations in quantum field theory and then study
in detail a very simplified model of the production of the quark-gluon
plasma-- Schwinger's mechanism for Pair-production from strong
``classical''  gauge fields. The value of doing a ``first principles''
calculation at this time, even if it is over-simplified, is multifold:
\item{(1)}  We can test the validity of existing  semiclassical transport
models of lepton production from the quark-gluon plasma.  We have already
discovered that these models have to be modified to correctly include Pauli
blocking and Bose enhancement effects which were ignored. \item{(2)}  We
can determine the effective hydrodynamics and show that certain kinematic
assumptions automatically lead to flat rapidity distributions independent
of the form of the equation of state.  \item{(3)}  We can determine the
dynamical equation of state and in the next order in a systematic
calculation in powers of (1/N) we will be able to  study whether
equilibration will occur and calculate self consistently lepton pair
production rates.

 First I would like to list the various approaches available to
study real time processes in Quantum Field Theory. Each of these approaches
needs an approximation scheme to reduce the number of degrees of freedom
in order to make the problem numerically tractable. Three methods that my
collaborators and I have studied in detail are:
\item{1-}  Functional Schrodinger Equation + variational
approximations \ref\rone{F. Cooper, S. Y. Pi and P. Stancioff, Phys. Rev. D.
34, 3831 (1986)}
\ref\r2{S. Y. Pi and M. Samiullah, Phys. Rev. D. 36 3128
(1987)}\ref\rthree{Fred Cooper and Emil Mottola, D. 36, 3114(1987)}
\item{2-}  Truncated Heisenberg Equations -- Large-N expansion or Mean-field
approximations to the Dyson equations
\ref\rfour{Fred Cooper and Emil Mottola, Phys. Rev. D 40, 456 (1989)}
\ref\rfive{F. Cooper, E. Mottola, B. Rogers, and P.
Anderson, in intermittency in High Energy Collisions, editied by F. Cooper,
R.C. Hwa and I. Sarcevic (World Scientific,
Singapore, 1991) p. 399}\ref\rsix{Y. Kluger, J. M. Eisenberg, B. Svetitsky, F.
Cooper, and E. Mottola, Phys. Rev. Lett. 67,
2427(1991)} \ref\r7{Y. Kluger, J. M. Eisenberg, B. Svetitsky, F. Cooper, and E.
Mottola, Phys. Rev. D 45, 4659 (1992)}.
\item{3-}  Schwinger's closed time path - Path Integral
Formalism in a large-N expansion \ref\reight{J. Schwinger, J. Math Phys. (N.Y.)
2, 407 (1961)}\ref\rnine{L.V. Keldysh, Sov. Phys.
JETP 20 1018 (1965)}\ref\rten{K. Chou, A. Su, B. Hao, L. Yu, Phys. Rpts. 118
(1985)}\ref\releven{R. D. Jordon, Phys. Rev. D. 33,
444 (1986)}.

There also exists an alternative formalism  related to the truncated
Heisenberg equations based on the Wigner Distribution function which has
been discussed by Rafelski and his collaborators \ref\r12{L.
Bialynicki-Birula, P. Gornicki, and J. Rafelski, Phys. Rev. D 44, 1825
(1991)}.  Once we have chosen a method we have to decide how to specify the
initial data at t=0. In these different approaches we have to specify
\item{1-} Initial position and width of say a Gaussian Wave
     Packet at t=0 in the Schrodinger picture
\item{2-}  Number and pair densities etc. in Heisenberg picture.
\item{3-}  Initial density matrix in the Path Integral Approach.

In classical field theory, such as classical electrodynamics, the theory is
finite and any smooth initial configuration of
the field is allowed  for the initial value problem.  When we have a
semiclassical field theory for the expectation value of
the fields however, the initial data can be reinterpreted in terms of the
particle language and even a smooth initial
configuration of the field might not be consistent with certain physical
constraints such as the initial state having finite
number density at t=0 with respect to an adiabatic vacuum.  (This requirement
is automatic for finite temperature field
theory).  Thus arbitrary initial data may not be consistent with
renormalizability. This is discussed in detail in \rthree \rfour.
    We also have an additional new problem to face -- how to perform
renormalization in a non-covariant formulation of the field theory.
To do this we  isolate the divergences in an adiabatic (WKB)  expansion of
Green's functions. This method is similar to the technique of adiabatic
regularization used by Parker and Fulling
\ref\rthirteen{L. Parker, and S. A. Fulling, Phys. Rev. D. 7 2357 (1973)} in
their study of  semiclassical gravity.
     The problem we will address in detail in these lectures is pair
production of either Bosons or Fermions from strong Classical Fields
which are either functions of time t,  or  fluid proper time $\tau=
(t^{2}-z^{2})^{1/2}$.
We will compare the results of the numerical simulation of this problem
(for the degradation of the field, the particle spectra, etc.)
with a semi-classical transport approach using a Schwinger-inspired
source term \ref\rfourteen{K. Kajantie and T. Matsui, Phys. Lett. 164B, 373
(1985)}\ref\rfifteen{G. Gatoff, A. K. Kerman, and T.
Matsui, Phys. Rev. D. 36, 114 (1987)}\ref\rsixteen{A. Bialas, W. Czyz, A.
Dyrek, and W.
Florkowski, Nucl. Phys. B296, 611 (1988)}.  We will also discuss the effective
hydrodynamics derived  from the expectation value of the energy momentum
tensor of the quantum theory.

\newsec{SUMMARY OF THE  DIFFERENT STRATEGIES IN $\lambda\varphi^{4}$ FIELD
THEORY}

For simplicity let us first study these different approaches to initial value
problems in the simplest case-
$\lambda\varphi^{4}$ field theory.
\item{a)} Schrodinger Picture:
     In the Schrodinger picture the Initial State is described by a wave
functional at t=0. For example a  Gaussian
wave functional is
\eqnn\nonumber
$$
\eqalignno{<\varphi | \Psi> &= \psi[\varphi, t] & \cr
&=\exp[-\int_{x,y}[\varphi(x)-\hat{\varphi}(x)]
[G^{-1}(x,y)/4-i\Sigma(x,y)][\varphi(y)-
\hat{\varphi}(y)]]  &(2.1)  \cr}
$$
The time evolution is given by the Functional Schrodinger equation \rone :
\eqnn\nonumber
$$
\eqalignno{i \partial\psi/\partial t &= H\psi & \cr
H &=\int d^{3}x [- {1 \over 2}
\delta^{2}/\delta \varphi^{2}+ {1 \over 2} (\nabla\varphi)^{2} + V(\varphi)] &
(2.2) \cr}
$$
This is a generalization of the usual Schrodinger equation:
\eqnn\nonumber
$$
\eqalignno{\psi(x) &= < \psi | x > , x\rightarrow \varphi(x,t); & \cr
p &= -i\delta/\delta x \rightarrow\pi = -i\delta/\delta\varphi & \cr
i\partial\psi/\partial t &=H\psi ; H=-\partial^{2}/\partial x^{2} + V(x) &
(2.3) \cr}
$$
with initial condition:
\eqn\twofour{\psi(0) = \exp [-\alpha(x-x_{o})^{2}].}

One might imagine solving (2.2) on a computer by introducing
a lattice in d dimensions and converting the functional
derivatives into partial derivatives.  One then quickly realizes
that the number of degrees of freedom in equation (2.2) is rather overwhelming.
To control this problem one uses variational trial
wave functionals which become ``exact'' in the large-N limit-- namely
Gaussians.
The equations of motion for the variational parameters can be obtained from
Dirac's variational principle \ref\r17{P. A. M. Dirac, Proc. Camb. Phil. Soc.
26 (1930)}:
\eqn\twofive{\Gamma = \int dt < \Psi | i \partial/\partial t -H| \Psi >}
$\delta\Gamma = 0 \rightarrow$  Schrodinger's equation:
\eqn\twosix{i \partial/\partial t -H|\Psi > = 0}

In the $\varphi$ representation one can choose a Gaussian
trial wave functional:
\eqn\twoseven{\eqalign{&<\varphi|\Psi_{v}>
=\psi_{v}[\varphi,t]=\exp[-\int_{x,y}[\varphi
(x)- \hat{\varphi}(x,t)]\cr
&[G^{-1}(x,y,t)/ 4-i\Sigma(x,y,t)][\varphi(y)-
\hat{\varphi} (y,t)]]+ i \hat{\pi}(x,t) [\varphi(x)- \hat{\varphi}(x,t)]\cr}}
where the variational parameters have the meaning:
\eqnn\nonumber
$$
\eqalignno{\hat{\varphi}(x,t) &= < \Psi_{v}|\varphi| \Psi_{v} >; \hat{\pi}(x,t)
=
<\Psi_{v}|-i \delta/\delta\varphi| \Psi_{v} > & \cr G(x,y,t) &= < \Psi_{v}|
\varphi(x)\varphi(y)| \Psi_{v} > - \hat{\varphi}(x,t) \hat{\varphi}(y,t) &
(2.8) \cr}
$$
Then the effective action for the trial wave functional is
\eqnn\nonumber
$$
\eqalignno{\Gamma(\hat{\varphi}, \hat{\pi},G,\Sigma) &=\int dt < \Psi_{v}| i
\partial/ \partial t -H| \Psi_{v}> & \cr
&= \int dt dx[\pi(x,t)\partial\varphi(x,t)/ \partial t +\int dt dx dy
\Sigma(x,y)\partial
G(x,y,t)/\partial t &\cr
&-\int dt < H > & (2.9) \cr}
$$
where
$$< H > = \int dx \{{\pi}^{2}/2 + 2 \Sigma G\Sigma + G^{-1}/8 + 1/2 (\nabla
\varphi)^{2} -1/2 \nabla^{2}G + 1/2 V''[\varphi] G + 1/8 V''''[\varphi]
G^{2}\}. $$

$< H >$ is a constant of the motion and is a first integral of the motion. For
$\lambda\varphi^{4}$ field theory we get the following equations of motion:
\eqnn\nonumber
$$
\eqalignno{\dot{\pi}(x,t) &= \nabla^{2}\varphi - \partial <V> /\partial\varphi;
& \cr
\dot{\varphi}(x,t) &= \pi & \cr
\dot{G}(x,t) &= 2\int dz[\Sigma (x,z)G(z,x)+G(x,z)\Sigma (z,x)] & \cr
\dot{\Sigma}(x,t) &= -2 \int dz[\Sigma (x,z)\Sigma (z,x) + G^{-2}/8 & \cr
& +  [{1 \over 2}\nabla^{2}_{x} - \partial <V> / \partial G] \delta^{3}(x-y) &
(2.10) \cr}
$$

If there is translational invariance and $\varphi$=0 we obtain  a second order
differential equation for G(k,t), the Fourier transform of G(x,t):
\eqnn\nonumber
$$
\eqalignno{& 2 \ddot{G} (k,t)G(k,t) - \dot{G}^{2}(k,t) +4 \Gamma
(k,t)G^{2}(k,t) - 1 = 0 & \cr
&\Gamma (k,t) = k^{2}+m^{2}(t); m^{2}(t) = \mu^{2} + {1 \over 2}\lambda \int
[dk] G(k,t) & (2.11) \cr}
$$

This approximation is called the time-dependent Hartree-Fock Approximation and
is equivalent to the leading term in a 1/N
expansion of the field theory \rthree . To understand this trial wave function
let us look at a simple quantum mechanics
problem- the  harmonic oscillator with a gaussian initial state. Harmonic
oscillator: V(x) =1/2 m x$^{2}$,

Initial conditions:
\eqnn\nonumber
$$
\eqalignno{\Psi(x,0) &= [2\pi G(0)]^{-1/2}\exp\lbrace-x^{2}/ [4 G(0)]\rbrace &
\cr
q(0) &= < x > = 0 & (2.12) \cr}
$$
For the harmonic oscillator a Gaussian remains Gaussian as time evolves so that
\eqn\twothirteen{\Psi(x,t) = (2\pi G(t))^{-1/2} \exp \lbrace -x^{2}[G^{-1}
(t)/4 - i \Sigma (t)]\rbrace }

We find that the conserved Energy can be written in terms of G as follows:
\eqn\twofourteen{E = < H > = \dot{G}^{2}/8G + Gm^{2}/ 2 + G^{-1}/8 =
\dot{G}^{2}/8G + V[g]}

We plot V[g] in fig 1.  From fig. 1 we see that the ground state is G = 1/(2m).
If at t=0,  G$_{0}$ = 1/(2M)  ; m $\not$ = M then
\eqn\twofifteen{G(t) = 1/2 (G_{0} + G_{1})  + 1/2 (G_{0}- G_{1}) \cos (2 m
(t-t_{o}))}

Thus the width oscillates with frequency 2m between G$_{0}$ and G$_{1}$.
Generalizing to free field theory (which is just independent harmonic
oscillators)  we have instead for each mode of momentum k:
\eqn\twosixteen{< H(k) > = \dot{G}^{2}/ 8G + (k^{2} + m^{2})G / 2 + G^{-1}/8}
This leads to the same result for G(k,t)  as for G(t) with m$\rightarrow
\omega_{k} = (k^{2} +m^{2})^{1/2}$.  However in field theory, unlike quantum
mechanics, an arbitrary initial Gaussian state is not necessarily a physically
valid choice since it might correspond to an infinite particle density or
energy
density when compared to the adiabatic vacuum. Thus the particle interpretation
implies that  one needs to restrict the large k behavior of G(k) at t=0 to be a
physically allowed initial state with finite particle number, energy density
etc.  Otherwise one gets extra unwanted infinities in loops.

\item{b)} Heisenberg Picture: Green's function approach

In problems where  there is spatial homogeneity  one
has a Fourier decomposition for  a charged  field $\varphi$ in terms of mode
functions $f_{k}(t)$ which depend only on the time and the usual creation and
annihilation operators a and b which satisfy the canonical commutation
relations:
\eqnn\nonumber
$$
\eqalignno{\Phi (x,t)&= \int [dk] [f_{k} (t) a_{k} e^{ikx} + f^{\ast} _{_k}(t)
b_{k}^{+} e ^{-ikx}] &
\cr [a_{k}, a^{+}_{k'}] &= [b_{k}, b^{+}_{k'}] = (2\pi)^{3} \delta^{3} (k-k') &
(2.17) \cr}
$$
 The initial state $|i >$ is totally specified by specifying at
t=0 the  matrix elements of a and b:
\eqnn\nonumber
$$
\eqalignno{< i| a^{+}_{k} a _{k} |i > &= (2\pi)^{d} \delta^{d} (k-k') n_{+}(k)
& \cr
< i| b_{k} a _{k} |i > &= (2\pi)^{d} \delta^{d} (k+k') F(k) etc. & (2.18) \cr}
$$

The equation for the expectation value of the equation of motion is:
\eqn\twonineteen{< i | ( -\mbox{0.1}{0.1} + m^{2})\varphi + \lambda
(\varphi^{+}\varphi) \varphi | i > = 0}

We see from these equations that we also need to solve the  equation of motion
for
$ < i | \lambda (\varphi^{+}\varphi) \varphi | i >$.

In general  we get a Heirarchy of  Green's function  equations- The BBGKY
heirarchy.
To make practical progress we need a truncation scheme which allows us to solve
the lowest order problem and then systematically calculate corrections. In the
large N expansion  the lowest order approximation leads to a factorization
\eqnn\nonumber
$$
\eqalignno{< i | (\varphi^{+}\varphi) \varphi | i > &= < i |
(\varphi^{+}\varphi)|i > < i| \varphi | i > & \cr &=G(x,x;t) < i| \varphi | i >
& (2.20) \cr}
$$
where the fourier transform G(k,t) of G(x-x'; t) obeys the same equation as the
width of the Gaussian wave
function in the Schrodinger equation in the Hartree approximation.
\eqnn\nonumber
$$
\eqalignno{&2 \ddot{G} (k,t)G(k,t) - \dot{G}^{2}(k,t) +4 \Gamma (k,t)G^{2}(k,t)
- 1 = 0 & \cr
&\Gamma(k,t) = k^{2}+m^{2}(t); m^{2}(t) = \mu^{2} + {1 \over 2}\lambda \int
[dk] G(k,t) & \cr
& G(x,x;t)=\int [dk] G(k,t) & (2.21) \cr}
$$
Thus the large-N expansion (Hartree approximation, mean field
approximation) truncates the hierarchy of coupled Green's function
equations making it necessary to only solve the coupled one and
two-point Green's function equations.

In these mean field equations the problem reduces to an external field problem
in that  the quantum field $\varphi$ obeys the equation:
\eqn\twotwentytwo{(-\mbox{0.1}{0.1} +m^{2}(t) )\varphi = 0}
Because we have an external field problem with spatial homogeneity:
the mode functions f(t) in (2.17) obey:
\eqn\twotwentythree{(\partial_{0}^{2} + \omega^{2})f=0; \;
\omega^{2}=k^{2}+m^{2}(t)}

The canonical commutation relations lead to a constraint on the mode functions:
\eqn\twotwentyfour{f_{k}\dot{f}_{k}^{\ast}-f_{k}^{\ast} \dot{f}_{k} = i}
which is automatically satified by the WKB form  ansatz:
\eqnn\nonumber
$$
\eqalignno{f_{k}(t) &= [2\Omega_{K}(t)]^{-1/2} \exp [-iy_{k}(t)] & \cr
\dot{y}_{k}(t) &= \Omega_{k}(t) & (2.25) \cr}
$$
which lead to the equation
\eqn\twotwentysix{\Omega_{k}^{2} (t) + \ddot{\Omega}_{k}/(2 \Omega_{k})
-{3 \over 4} (\dot{\Omega}_{k}/\Omega_{k})^{2} = \omega_{k}^{2}(t).}
At  t=0 one has in general for the initial state:

$$
\eqalignno{& < i|a^{+}_{k} a_{k}|i > = (2\pi)^{d} \delta^{d} (k-k') n_{+}(k) &
\cr
& < i| b_{k} a_{k}|i > = (2\pi)^{d} \delta^{d} (k-k') F(k) & \cr }
$$

For an adiabatic vacuum: n(k) =F(k) = 0, and the initial conditions
on $\Omega$ are
\eqn\twotwentyseven{\Omega(k,t=0) =\omega(k,t=0); \dot{\Omega}(k,t=0) =
\dot{\omega}(k,t=0).}

This formalism, however is perfectly general and one could
take any initial state with an integrable phase space particle
density n(k) and pair density F(k). As a particular choice
one could  have chosen at t=0 an equilibrium
configuration of pions described by a temperature  kT = $\beta^{-1}$
\eqn\twotwentyeight{n(k) = 1/(\exp[\beta E(k)]-1)}
\item{c)} Path Integral Approach: Closed time-path formalism

The only formalism that allows a  systematic approach to initial value problems
is the closed time-path approach of J. Schwinger\reight which  was further
elaborated by
Keldysh\rnine and put into a Path Integral framework by Chou, Su, Hao and
Yu\rten . This Path Integral
approach allows standard Path Integral approximation schemes such as the large
N approximation as well as ensuring
causality for the Green's functions for initial value problems
\ref\rtwentyeight{P. Anderson, F. Cooper, S. Habib, E. Mottola, and J. Paz [in
preparation]}. The starting point for determining the Green's
functions of the initial value problem is the generating Functional:
\eqn\twotwentynine{Z [J^{+}, J^{-}, \rho] = < i| T^{\ast}(\exp \{ - \int
iJ_{-}\varphi_{-}\} )|out > < out | T(\exp \int
iJ_{+}\varphi_{+})| i >}

This can be written as the product of an ordinary Path integral times a complex
conjugate one or as a matrix Path
integral.
\eqnn\nonumber
$$
\eqalignno{Z [J^{+}, J^{-}, \rho] &=\int d\varphi^{+} d\varphi^{-} <
\varphi_{+}, i| \rho |\varphi_{-}, i > \exp i[(S[\varphi_{+}] +
J_{+}\varphi_{+}) -
(S^{\ast}[\varphi_{-}] +J_{-}\varphi_{-}) ] & \cr
&= \int d\varphi_{\alpha} \exp i (S[\varphi_{\alpha}] +
J_{\alpha}\varphi_{\alpha} ) <
\varphi_{1},i|\rho |\varphi_{2} , i > & (2.30) \cr}
$$
where   $< \varphi_{+} i| \rho|\varphi_{-}, i >$ is the density matrix defining
the
initial state.

This leads to the following matrix Green's functions \releven :
\eqnn\nonumber
$$
\eqalignno{G_{++}&= \delta^{2}\ln \; Z/\delta J^{+} \delta J^{+}|_{j=0}= <
T(\varphi (x_{1}), \varphi (x_{2}) > & \cr
G_{--}&=\delta^{2}\ln \; Z/\delta J^{-} \delta J^{-}|_{j=0}= <
T^{\ast}(\varphi(x_{1}), \varphi(x_{2}) > & \cr
G_{+-}&= \delta^{2}\ln \; Z/\delta J^{+} \delta J^{-}|_{j=0} = <
\varphi(x_{2}), \varphi(x_{1}) > & \cr
G_{-+}&= \delta^{2}\ln \; Z/\delta J^{-} \delta J^{+}|_{j=0} = <
\varphi(x_{1}), \varphi(x_{2}) > & (2.31) \cr}
$$

The matrix Green's function structure insures causality. In this approach it is
easy to generate a 1/N expansion in analogy with ordinary
field theory. The diagrams are the same as in the usual 1/N expansion, except
the Green's functions are the matrix Green's functions described
above.If in lowest order in (1/N) we have an external field problem as
described above, one can directly use the mode solutions of the
previous methods to determine the lowest order matrix Green's function of eq.
(2.31).  This obviates the need to discuss the initial density
matrix of the theory, since it is these Green's functions which then enter the
diagrams of the higher order calculations.

\newsec{MAIN EXPANSION IDEA: FLAVOR SU(N)}

In many problems one of the fields can be treated classically to first
approximation-- pair production in Strong Electric or
Gravitational fields. This makes the lowest order problem an external field
problem.  One way to generate a systematic
expansion whose lowest order is an external field problem
is by introducing N copies of the original problem and expanding in Flavor
SU(N).  This is most easily done in the Path Integral formalism.
 For the initial value problem one would use the matrix Green's functions
discussed above. Having an extra large parameter N allows an
evaluation of the Path integral by Laplace's method (or the method of Steepest
Descent). To obtain the large N expansion one realizes that if
there are N flavors the loops carry an extra N. Rescaling the fields then
display an overall factor of N in the effective action which
includes the loops.
Examples:
\eqnn\nonumber
$$
\eqalignno{\lambda \varphi^{4}:  \chi & = \varphi^{2} & \cr
Z &= \int d\chi\int d\varphi \exp [-\int
(\partial_{\mu}\varphi)^{2}+\lambda\chi\varphi^{2} -\lambda\chi^{2}
+\mu^{2}\varphi+J\varphi+S\chi] & \cr
\varphi\rightarrow\varphi_{i}, i & =1,2,\cdots  N;
\lambda\rightarrow\lambda/N; \varphi_{i}\rightarrow N^{1/2}\varphi_{i};
\chi\rightarrow N\chi , \lambda\chi\rightarrow\lambda \chi .} $$
Integrating over $\varphi$ we obtain:
\eqnn\nonumber
$$
\eqalignno{& Z=\int d\chi \exp \lbrace - N [\chi^{2} +{1 \over 2} Tr \ln
G^{-1} -jGj]\rbrace & \cr & = \int d\chi \exp \lbrace -N
S_{eff}(\chi)\rbrace & \cr & G^{-1} = [- \mbox{0.1}{0.1} +\mu^{2}
+\lambda\chi ] \delta (x-y) & (3.1) \cr} $$

Evaluating the Path Integral  at the Saddle point, $\delta
S_{eff}(\chi)/\delta\chi =0$ leads to the self consistent external field
problem

\eqn\threetwo{[- \mbox{0.1}{0.1}  + \mu^{2} +\lambda\chi ] \varphi =0; \chi
=\varphi^{2} + G(xx)}

In QED we obtain an external field problem by integrating out the
fermions (which have now N flavors to give an extra N to
the determinant) and then rescaling the fields to display the
overall factor of N in the effective Action:
QED:
\eqnn\nonumber
$$
\eqalignno{Z & =\int d A_{\mu} \int d\overline{\Psi} d\Psi\exp [\int
dx\lbrace-{1 \over 4} F^{2} + \overline{\Psi}(i\gamma \partial - e \not A +
m)\Psi \rbrace + \overline{\Psi}\eta +\overline{\eta}\Psi]\cr
&\Psi\rightarrow \Psi_{i}; e\rightarrow e/\sqrt{N}, A\rightarrow A
\sqrt{N} & (3.3) \cr}
$$

Integrate out the N species of fermions
\eqnn\nonumber
$$
\eqalignno{& \int d A_{\mu} \exp\lbrace-N S_{eff}(A_{\mu})\rbrace & \cr
& S_{eff}(A_{\mu}) = \int dx {1 \over 4} F^{2} + Tr \ln (S^{-1}(x,y; A)) +
\overline{\eta}S(x,y; A)\eta ] & \cr
& S^{-1}(x,y; A) = ( i\gamma\partial -e \not A (x) + m)\delta (x-y) & (3.4)\cr}
$$

Evaluating the Path Integral at the saddle point, $\delta
S_{eff}(A_{\mu})/\delta A_{\mu} =0$ leads to the external field problem:
\eqn\threefive{(i\gamma\partial -e\not{A}+m) \Psi = 0}
where $A$ is an external field,$\Psi$ is a quantum field.  We also obtain the
semiclassical Maxwell Equation:
\eqn\threesix{\partial_{\mu} F^{\mu\lambda}=< j^{\lambda} > = e
<\overline{\Psi}\gamma^{\lambda}\Psi >.}

In all these problems one has in leading order in 1/N a straightforward problem
of a quantum field theory in a background
field which allows a normal mode decomposition in terms of the solutions of the
classical field equations.   Renormalization can be
carried out by an adiabatic expansion of the mode equation\rthirteen .  The
effect of quantum fluctuations about the semiclassical
field can be systematically taken into account by calculating the fluctuations
about the leading stationary phase point in
the Path Integral order by order in the 1/N expansion.

\newsec{PARTICLE PRODUCTION IN THE CENTRAL RAPIDITY REGION IN HEAVY ION
COLLISIONS}

   A popular picture of high-energy heavy ion collisions begins with the
creation of a flux tube containing a strong color electric
field\ref\reighteen{S. Nussinov, Phys. Rev. Lett. 34, 1296 (1975)}.
The field energy is converted into particles as $ q\overline
{q}$ pairs and gluons which are created by tunnelling- the so-called Schwinger
mechanism \ref\rnineteen{J. Schwinger, Phys. Rev
82, 664 (1951)}\ref\r20{C. Itzykson and J. Zuber Quantum Field Theory.
McGraw-Hill (1980)}\ref\r21{A. Casher, H. Neuberger,
S. Nussinov, Phys Rev. D. 28, 179 (1979)}.  The particle production can be
modeled as an inside-outside cascade which is
symmetric under longitudinal boosts and thus produces a plateau in the particle
rapidity distribution.  The boost invariant
dynamics, in a hydrodynamical picture gets translated into energy densities
(such as E$^{2}$ )  being functions of the
proper time. We take this as an initial condition on the fields in an initial
value problem  based on this pair-production
mechanism. First let us look at the case where the electric field is a function
of real time t, treating later the more
realistic case where E= E($\tau); \tau = (t^{2} - z^{2})^{1/2}$. Thus we first
consider particle production from a spatially
uniform electric field such as that produced between two parallel plates. This
is an idealized model of a flux tube for QCD.
The problem of pair production from a constant Electric field (ignoring the
back reaction) was studied by J. Schwinger in
1951 \rnineteen . The physics is as follows:
    One imagines an electron bound by a potential well of order
 $|V_{0} |\approx$ 2m and submitted to an additional electric potential
eEx (as shown in  fig. 2 ).  The ionization probability  is proportional to the
WKB barrier penetration factor:

\eqn\fourone{\exp [-2 \int_{o}^{V_{o}/e} dx \lbrace 2m (V_{o} -|eE|
x)\rbrace^{1/2}]
= \exp (-{4 \over 3} m^{2}/|eE|)}

A direct calculation due to Schwinger from first principles using the effective
action in an arbitrary constant electric field (ignoring the back reaction)
gives instead
\eqn\fourtwo{w = [\alpha E^{2}/(2\pi^{2})] {\Sigma_{n=1}^{\infty}} {(-1)^{n+1}
\over n^{2}} \exp (-n\pi m^{2}/|eE| ).}

This equation tells us that pair production is exponentially suppressed unless
eE $\geq \pi m^{2}$.  So we expect (as we find in fig.
3) that there is a crossover  value of E where the  time it takes for E to
first reach zero (remember there are plasma oscillations)
is  relatively short.     Schwinger's result only applies when we can ignore
dynamical photons (as well as back reaction)and is related
to the lowest order in 1/N calculation where the electric field is treated as a
classical object. Schwinger's analytical result was
subsequently used as source term for an approximate transport theory
\rfourteen, \rfifteen, \rsixteen  approach to the  back reaction
connected with pair production which we will later compare with our exact
numerical results.

We will choose the electric field in the z direction and choose a particularly
simple gauge:

\eqn\fourthree{{\buildrel\rightarrow\over E} = E(t) \hat{k};
{\buildrel\rightarrow\over A} = A(t)\hat{k}; E(t)  = -dA/dt}

To maintain spatial homogeneity we have from Maxwell's equation:

\eqn\fourfour{\nabla^{.} E=\rho} that the plasma of  produced particles must be
neutral.
In scalar QED, the equation for the quantum field $\varphi$ is
\eqn\fourfive{-(\partial_{\alpha} -ieA_{\alpha}) (\partial
^{\alpha}-ieA^{\alpha}) \Phi + \mu^{2}\Phi = 0}
and for the electromagnetic field:
\eqn\foursix{\partial_{\alpha}F^{\beta \alpha} = < {\cal C} \lbrace - ie
(\Phi^{+}\partial^{\beta}\Phi - \Phi
\partial^{\beta}\Phi^{+}) - 2e^{2} A^{\beta}\Phi^{+}\Phi\rbrace >}
where ${\cal C}$ denotes charge symmetrization with respect to $\Phi^{+}$ and
$\Phi$.
For our constraints on the field E and our choice of gauge we get:
$$\eqnn\fourseven\eqalignno{-dE/dt & = < j_{Z} > = e \int [dk] (k_{Z}-eA (t) )
G(k,t)  & (4.7a)  \cr
{\rm where}\; G(k,t) & = [ < \varphi^{\dagger} \varphi + \varphi
\varphi^{\dagger} > -2 \varphi^{\ast}\varphi] _{FT}& (4.7b) \cr}
$$
For QED we have instead the field equation:
\eqn\foureight{[i \gamma \partial - e {\not A}(t) - m ] \Psi(x,t) = 0}
and the semiclassical Maxwell equation:
\eqn\fournine{- dE/dt = < j_{Z} > = {1\over 2} e < i|
[\overline{\Psi}(x,t),\gamma_{3}\Psi(x,t)]| i >}

The fact that the external field is independent of space (spatial homogeneity)
means that  one has a simple normal mode expansion
 of the fields just as in $\lambda\varphi^{4}$ field theory described earlier.

For Scalar QED we have
\eqnn\nonumber
$$
\eqalignno{\Phi(x,t) &= \int [dk] [f_{k} (t) a_{k} e^{ikx} + f^{\ast} _{_k}(t)
b_{k}^{+} e^{-ikx}] & \cr [\partial_{o}^{2} +
\omega_{k}^{2} (t) ]  f_{k} (t) &= 0 & \cr
\omega_{k}^{2} (t) &= [k-eA(t) ]^{2} + \mu^{2} + k_{\bot}^{2} & (4.10)  \cr}
$$

Repeating the arguments of (2.22 - 2.25) we again obtain for the
generalized frequency $\Omega_{k}$(t):
\eqn\foureleven{\Omega_{k}^{2}(t) + \ddot{\Omega}_{k}/(2 \Omega_{k}) -{3 \over
4} (\dot{\Omega}_{k}/\Omega_{k})^{2} =
\omega_{k}^{2}(t).} where now $\omega$ is given by (3.10) Spatial homogeneity
requires  translational invariance,
$$W(x-x', t,t') = \int [dk] W(k,t,t') e^{i k (x-x')}.$$

This in turn requires that
\eqnn\nonumber
$$
\eqalignno{< a^{+}_{k} a_{k} > &= (2\pi)^{d} \delta^{d} (k-k') n_{+}(k) & \cr
<  b^{+}_{k} b_{k} > &= (2\pi)^{d} \delta^{d} (k-k') n_{-}(k); & \cr
< b_{k} a_{k} > &= (2\pi)^{d} \delta^{d} (k+k') F(k) & (4.12) \cr}
$$

Thus we obtain for G(k,t)
\eqn\fourthirteen{G(k;t) = \Omega^{-1}(k,t) \lbrace 1+n_{+}(k) + n_{-}(k) + 2
F(k)\cos [2y_{k} (t)]\rbrace}

This is the most general form of the propagator that one would
use in the diagrams of the 1/N expansion, where n and F are the
particle and pair phase space densities at t=0. These parameters
also totally specify (in leading order in 1/N) the density matrix
at t=0.  To solve the field theory in leading order in 1/N (ignoring questions
of renormalization to be discussed below)
one solves the second order differential equation for each mode function
$\Omega_{k}$(t), determines G(k,t) and then solves the back reaction
equation:
\eqn\fourfourteen{-dE/dt = e \int [dk] (k_{Z}-eA (t) ) G(k,t)}

For QED one has to deal with the spinor structure:
\eqn\fourfifteen{\psi(x,t) = \int [dk] [u_{ks} (t) b_{k} e^{ikx} + v_{-ks}(t)
d_{-k}^{\dagger} e^{-ikx}]}

If we choose a basis where $\gamma^{0}\gamma^{3}$ is diagonal:
\eqnn\nonumber
$$
\eqalignno{\gamma^{0}\gamma^{3} \chi_{s} &= \lambda_{s}\chi_{s}, s=1, 2
\rightarrow\lambda=1; s=3,4 \rightarrow \lambda=-1 &\cr
\chi^{\dagger}_{r}\chi_{s} &= 2 \delta_{rs} & (4.16) \cr}
$$

Then the spinors u and v obey the equation
\eqn\threeseventeen{\lbrace \gamma^{0}\partial_{t} + i \gamma^{3}\pi +
i\gamma^{\bot}p^{\bot} + m\rbrace \pmatrix{u_{ks} & (t)\cr
v_{ks} & (t) \cr} =0}

Squaring the Dirac equation by letting:
\eqn\threeeighteen{\pmatrix{u_{ks}\cr
v_{ks}\cr}= \lbrace -\gamma^{0} \partial_{t} - i \gamma^{3}\pi - i
\gamma^{\bot}p^{\bot} +
m\rbrace \pmatrix{\chi_{s}f_{k}^{+}(t)\cr \chi_{s} f_{-k}^{-}(t)\cr}}
we find that the mode functions $f$ now obey:
\eqnn\nonumber
$$
\eqalignno{[\partial_{0}^{2} + \omega_{k}^{2} (t) -i \lambda_{s}\dot{\pi} ]
f_{k} (t) &= 0, & \cr \omega_{k}^{2} (t) &= \pi^{2}
 +p_{\bot}^{2}+ \mu^{2} & \cr \pi&=k-eA & (4.19)  \cr}
$$

If the operators $a_{k}$ and $b_{k}$ obey the usual anticommutation relations:

\eqn\fourtwenty{\lbrace a_{ks}, a^{\dagger}_{k's}\rbrace = \lbrace b_{ks},
a^{\dagger}_{k's}\rbrace = (2 \pi)^{3} \delta^{3} (k-k')\delta_{ss'}}
the $f_{k}$ are constrained to satisfy
\eqn\fourtwentyone{\omega^{2} f^{\ast\alpha} f^{\beta} + \dot{f}^{\ast\alpha}
\dot{f}^{\beta}
+i\lambda_{s}\pi [f^{\ast\alpha}\dot{f}^{\beta}-\dot{f}^{\ast\alpha}f^{\beta}]
=
\delta^{\alpha \beta}/2} Parametrizing the positive and negative frequency
solutions: \eqn\fourtwentytwo{f_{\pm}(t) =N_{\pm} \exp
\int_{0}^{t}g_{\pm}(\tau)d\tau,} we find:
\eqn\fourtwentythree{g^{+} = - [\lambda_{s}\dot{\pi}+ \dot{\Omega}]/2\Omega - i
\Omega}
where the generalized frequencies, $\Omega_{k}(t)$ now satisfies the equation:
\eqn\fourtwentyfour{\Omega_{k}^{2} (t) + \ddot{\Omega}_{k}/(2 \Omega_{k}) -{3
\over 4}
(\dot{\Omega}_{k}/\Omega_{k})^{2}-\dot{\pi}^{2}/(4\Omega^{2})- \lambda_{s}
\dot{\pi}\dot{\Omega}/\Omega^{2}}

Ignoring renormalization, the solution of QED is obtained by simutaneously
solving for these modes and also for E(t) which is obtained from the Maxwell
equation:
\eqn\fourtwentyfive{dE/dt = 2e\Sigma_{s=1}^{4} \int [dk] (p_{\bot}^{2}+m^{2})
\lambda_{s}| f^{+}_{ks}(t)|^{2}}

\newsec{RENORMALIZATION}

The equations of the previous section as they stand are not finite in the
continuum since the sum over modes in (4.14) and (4.25) contains a
divergence related to the renormalization of the charge (as well as the wave
function) resulting from the charged particle loops in the
definition of the current.

Let us first look at Scalar QED where the back-reaction equation is:

\eqn\fiveone{ -dE/dt = < j > =e \int [dk] (k_{z}-eA (t)) \Omega^{-1} [ 1+
N(k)...]}

We first see that N(k) has to fall fast enough at large k to not lead to any
further divergences--
this is equivalent to the condition that  the initial number density $\rho$ is
finite. The integral of $\Omega^{-1}$
contains a divergence proportional to dE/dt which renormalizes the charge (as
well as the field E).  To isolate this
divergence one makes an adiabatic expansion of the equation for the generalized
frequencies $\Omega_{k}$. That is, we imagine that the time
derivatives are small d/dt$\rightarrow \epsilon$ d/dt :

\eqn\fivetwo{\epsilon^{2}[\ddot{\Omega}_{k}/ (2 \Omega_{k}) -{3 \over 4}
(\dot{\Omega}_{k}/\Omega_{k})^{2}]=\omega_{k}^{2}(t)-\Omega_{k}^{2}(t)} and we
then expand in powers of $\epsilon
$
\eqn\fivethree{1/\Omega_{k}=1/\omega_{k} [1 +
\epsilon^{2}\lbrace\ddot{\omega}_{k}/4 \omega_{k} -{3 \over 8}(\dot
{\omega}_{k}/\omega_{k})^{2}\rbrace + 0 (\epsilon^{4} \omega_{k}^{-4})]}

We see that terms with higher derivatives are associated with more
convergence factors of 1/k so that one only has to consider the first
two terms in the adiabatic expansion to isolate the divergences which
are interpreted as the standard charge renormalization.
The log divergence comes from the term
\eqn\fivefour{\ddot{\omega}_{k} = e (dE/dt) (k-eA)\omega^{-1}}
After integrating over k this leads to a term of the form:
\eqn\fivefive{e^{2}\delta e^{2} dE/dt ; \delta e^{2} = {1\over 12} \int
[dk] \omega_{k}^{-3} = \pi(0)} where $\pi$(0) is the usual vacuum
polarization at
 q$^{2}$=0.  Subtracting this term from both sides of eq.
(5.1) we obtain:
\eqn\fivenine{-e dE/dt(1 +e^{2}\pi (0)) =
e^{2}[\int [dk](k_{z}-eA (t)) G -e\pi (0)dE/dt].}
 The Ward identity tells us that $eE =e_{R}E_{R}$; and the
renormalized charge is determined by
\eqn\fivesix{e_{R}^{2} = e^{2}/( 1+e^{2}\pi (0))} so the
explicity mode by mode finite renormalized equation is
\eqn\fiveseven{- dE_{R}/dt = e_{R}\int [dk] (k-eA
(t))[\Omega^{-1}-\omega^{-1} -e_{R}^{2} (k-eA
(t))(dE/dt)\omega^{-5}/4]}

For QED one gets instead after charge renormalization:
\eqn\fiveeight{dE_{R}/dt = 2e_{R} \Sigma_{s=1}^{4}\int[dk] [(p_{\bot}^{2}
+m^{2})\lambda_{s}| f^{+}_{ks}|^{2} -e_{R}^{2}dE_{R}/dt \; \omega^{-3}}

\newsec{HEAVY ION COLLISIONS AND BOOST INVARIANT DYNAMICS}

  In e$^{+}$ e$^{-}$ annihilation, hadronic collisions and in heavy-ion
collisions
particle production in the central rapidity region can be modeled as an
inside-outside cascade which is symmetric under longitudinal boosts which leads
to a plateau
in the particle rapidity distributions. This boost invariance also emerges
dynamically in Landau's hydrodynamical model \ref\rtwentytwo{L. D.
Landau, Izv. Akad. Nauyk SSSR 17 (1953) 51} and forms an essential kinematic
ingredient in the analyses of Cooper, Frye and Schonberg
\ref\rtwentythree{F. Cooper, G. Frye and E. Schonberg, Phys. Rev. D 11(1975)
192} as well as Bjorken\ref\r24{J. D. Bjorken, Phys. Rev. D 27
(1983), 140.}. It was recognized by Cooper and collaborators  and further
elaborated by Bjorken that in a hydrodynamical framework
scale invariant initial conditions :
\eqn\sixone{v=z/t , \epsilon(x,t) \rightarrow \epsilon (\tau),
\tau^{2}=t^{2}-z^{2}}
would automatically lead to flat rapidity distributions.  In the context of
transport or field theory modelling of the heavy ion
collision, after an initial time $\tau_{0}$, energy densities are expected to
be functions only of the  fluid ``proper time'' $\tau$.  We therefore
assume that the kinematics makes the electric field E only a function of the
proper time $\tau$. For this kinematical choice it is convenient
to introduce new variables $\tau, \eta$  the fluid ``proper time'' and the
fluid rapidity (when v=z/t) via :
\eqn\sixtwo{z=\tau\sinh\eta,t=\tau\cosh\eta .}
This change of coordinates to $(\tau,\eta)$ from (t,z) can be accommodated by
the usual formalism
of curved space \ref\rtwentyfive{N. D. Birell and P.C. W. Davies  Quantum
Fields in Curved Space (Cambridge University press, Cambridge,
England, 1982)}\ref\rtwentysix{S. Weinberg, Gravitation and Cosmology:
Principles and Applications of the General Theory of Relativity(Wiley,
New York, 1972)} (except the curvature here is zero).  One introduces the
metric in curved space
\eqn\sixthree{g_{\alpha \beta} = {\rm diag}(-1, 0, 0, \tau^{2}).}
Maxwell's equations
\eqn\sixfour{(-g)^{-1/2}\partial_{\beta} [ (-g) ^{-1/2} F^{\alpha \beta}] =
j^{\alpha}}
becomes for an electric field $E(\tau)$ in the z direction
\eqn\sixfive{E(\tau) = F_{zt}=F_{\eta\tau}/\tau = -\tau^{-1}\partial_{\tau}
A(\tau)}
\eqn\sixsix{-1/\tau \partial_{\tau} [1/\tau \partial_{\tau} A (\tau) ] = <
j^{\eta} > }
For Scalar Electrodynamics the equation for $\chi = \sqrt{\tau} \varphi$ is
\eqn\sixseven{(\partial^{2}_{\tau} + \tau^{-2} [ (\partial_{\eta}-ieA(\tau)
)^{2} + 1/4] -
\partial_{x}^{2}-\partial_{y}^{2} +m^{2})\chi =0}

The rescaled field $\chi$ has the same Fourier decomposition as $\phi$ had in
flat space with the mode functions f obeying
\eqn\sixeight{[\partial_{\tau}^{2} + \omega_{k}^{2} (\tau) ] f_{k} (\tau) = 0}
however now
\eqn\sixnine{\omega_{k}^{2}(\tau) = [k-eA(\tau) ]^{2}/\tau^{2} + k_{\bot}^{2} +
\mu^{2} + 1/(4\tau^{2})}
so that the longitudinal momenta get suppressed at large $\tau$.  For fermions
one has the added complication that the
covariant derivative now has a spin piece:  (denote the Minkowski indices with
$\alpha, \beta$ the curvilinear coordinates with
$\mu \nu$)
\eqnn\nonumber
$$
\eqalignno{\nabla _{\mu} &= \partial_{\mu} + \Gamma_{\mu} -ieA_{\mu} & \cr
\Gamma_{\mu} &= {1\over 2} \Sigma^{\alpha \beta} V_{\alpha}^{\upsilon}
V_{\beta\upsilon;\mu} & \cr
\Sigma^{\alpha \beta} &= {1\over 4} [\gamma^{\alpha}, \gamma^{\beta}] & (6.10)
\cr}
$$
and the vierbein represents the transformation to the Minkowski coordinates:
\eqn\sixeleven{g_{\mu\nu} = V_{\mu}^{\alpha}V_{\nu}^{\beta} \eta_{\alpha
\beta};\; {\buildrel\sim \over \gamma}^{\mu} =
\gamma^{\alpha}V_{\alpha}^{\mu}}

Maxwell's equation becomes:
\eqnn\nonumber
$$
\eqalignno{-\tau^{-1} dE(\tau )/ d\tau = < j^{\eta} > & = {1\over 2} e < i|
[\overline{\Psi }, {\buildrel\sim \over \gamma}^{\eta }\Psi ]|i > & \cr & =
{1\over 2\tau} e < i| [\Psi^{\dagger}, \gamma^{0}\gamma^{3}\Psi ]| i > &
(6.12) \cr } $$ and the fermion mode functions now obey

\eqnn\nonumber
$$
\eqalignno{[\partial_{\tau}^{2} + \omega_{k}^{2} (\tau) - i
\lambda_{s}\dot{\pi} ] f_{k} (\tau) &= 0 &
(6.13)  \cr} $$
where
\eqn\sixthirteenb{\omega_{k}^{2} (\tau) = \pi^{2} +p_{\bot}^{2} + \mu^{2};
\pi = (p-eA (\tau))/\tau}

The divergences in Maxwell's equation in curved space can be
 renormalized as before by an adiabatic expansion in the variable $\tau$.
The details of this calculation are presented in \ref\rtwentynine{F. Cooper, J.
M. Eisenberg, Y. Kluger, E. Mottola and B. Svetitsky
``Particle Production in the Central Rapidity Region'' Tel-Aviv Preprint TAUP
1944-92.}:

\newsec{PARTICLE PRODUCTION RATES AND THE BOGOLIUBOV TRANSFORMATION}

The wave functions of the first order adiabatic expansion  e$^{ikx}
f^{0}_{k}(t)$ where
\eqn\sevenone{f^{0}_{k}(t)= (2\omega_{k})^{-1/2}\exp [-i
\int_{0}^{t}\omega_{k}(t') dt']} form an alternative basis for expanding the
scalar fields and allows one to define an interpolating number density N(k,t)
which becomes the true one as t$\rightarrow \infty$. Expanding the field in
terms of f$^{0}_{k}$(t) we have

\eqn\seventwo{\Phi(x,t)=\int [dk][a_{k}^{0}(t)e^{ikx}f^{0}_{k}(t)+ f^{\ast
0}_{-k}(t) b_{k}^{0\dagger}(t) e^{-ikx}]
{\rm where}\; a_{k}^{0}(t\rightarrow\infty) = a_{k}^{out}\; etc.}
In this expansion the creation and annihiliation operators are time dependent.
We also have our previous expansion in
terms of the time independent operators a and b related to the initial state:
\eqn\seventhree{\Phi(x,t) = \int [dk] [f_{k} (t) a_{k} e^{ikx} + f^{\ast}_{-k}
(t) b_{k}^{\dagger}
e^{-ikx}].}

  We recognize that $a_{k}$ and $a_{k}^{0}(t)$ are related by a unitary
transformation.
The Bogoliubov coefficients are defined by
\eqnn\nonumber
$$
\eqalignno{a_{k}^{0}(t) &= \alpha (k,t)a_{k} + \beta^{\ast} (k,t)
b_{k}^{\dagger} & \cr
b_{k}^{0}(t) &= \alpha(k,t) b_{k} +\beta^{\ast} (k,t) a_{k}^{\dagger} & \cr
|\alpha(k,t)|^{2} + |\beta (k,t)|^{2} &= 1 & (7.4) \cr} $$
The number of particles produced per unit volume is just
\eqn\sevenfive{V^{-1} dN/dk =  < t=0 | b_{k}^{out \;\dagger} b_{k}^{out} +
a_{k}^{out \; \dagger} a_{k}^{out} | t=0 >}

The interpolating number density is defined in terms of the first order
adiabatic operators:
\eqnn\nonumber
$$
\eqalignno{V^{-1} dN(k,t) /d^{3}k &= \langle t=0 |b_{k}^{0 \; \dagger}(t)
b_{k}^{0} (t) + a_{k}^{0 \; \dagger}
(t)a_{k}^{0} (t) | t=0\rangle & \cr
&=(1+N(k))|\beta |^{2} + N(k)|\alpha|^{2} + 2 Re \lbrace\alpha \beta
F(k)\rbrace & (7.6) \cr} $$

For N=F=O (the adiabatic vacuum at t=0)
\eqn\sevenseven{V^{-1} dN(k,t) /dk =|\beta|^{2}= (4\omega_{k}\Omega_{k})^{-1} [
(\Omega_{k}-\omega_{k})^{2} + {1 \over 4}
(\dot{\Omega}_{k}/\Omega_{k} - \dot{\omega}_{k}/\omega_{k})^{2} ]} We see that
adiabatic initial conditions (no particle
production at t=0) are

\eqn\sixeight{\Omega_{k}(0)=\omega_{k}; \dot{\Omega}_{k}(0) =
\dot{\omega}_{k}(0)}

For fermions we have instead:

$$
{V^{-1} dN(k,t) /dk = \Sigma_{s}w(\omega +\lambda \pi)(2\omega)^{-1}
[\omega^{2}|f^{+}|^{2} +|\dot{f}^{+}|^{2} - i
\omega (f^{\ast +} \partial_{0}f^{+} - f^{+} \partial_{0}f^{\ast +})].}
$$

Similar expressions exist for the boost invariant problem \rtwentynine .

\newsec{TRANSPORT APPROACH TO MULTIPARTICLE PRODUCTION}

A classical kinetic theory approach to the back-reaction problem as discussed
in \rfourteen \rfifteen \rsixteen introduces a phase space
single particle distribution function $f(x,p,t)$ in the presence of a
homogeneous electric field and with a phenomenological source term inspired by
Schwinger's solution for the constant external field.  \eqnn\nonumber
$$
\eqalignno{df/dt &= \partial f/\partial t +e E(t) \partial f/\partial p & \cr
&=dN/dt dz dp & \cr
&=|e E(t)|\ln [1\pm \exp[-\pi
m_{\bot}^{2}/ |e E(t) |]]  \delta (p) &(8.1) \cr}
$$
$\pm$ stand for boson(+) or fermion case (-).
The right hand side of (8.1) is a naive use of  Schwinger's formula (valid when
no particles are present and for constant fields
 with E replaced by  E(t)  $\lbrace$or E($\tau)\rbrace$. This approach was
recently used to predict dilepton production from the
quark-gluon plasma \ref\rtwentyseven{M. Asakawa and T. Matsui, Phys. Rev. D43,
2871 (1991)}.
A potential problem with replacing constant E by E(t)
 is that in the field theory simulations E(t) is rapidly varying in time. A
more serious problem is that once particles are produced, Schwinger's
derivation, which was for particle production from the vacuum, is no
longer valid. This however can be fixed up by the following arguement.
Once particles are present there is an additional quantum mechanical effect due
to statistics-- Bose enhancement or Pauli Suppression.  For the
external field problem one always has a normal mode decomposition at each time
t. Thus the creation and annihilation operators at different t
are again connected by a unitary transformation:

$$
\eqalignno{b(k, t+\Delta t) &= \alpha(t + \Delta t) b(k,t) + \beta (t+\Delta t)
d^{\dagger}(k,t) & \cr
|\alpha|^{2} +|\beta |^{2} &= 1; |b^{\dagger} b| = n_{+}; |a^{\dagger} a| =
n_{-}; n_{+} = n_{-} = n & (8.2) \cr}
$$
Therefore
\eqn\eightthree{n(t+\Delta t) = n(t) + 2 |\beta(t+\Delta t)|^{2} \lbrace 1 \pm
n\rbrace}
or
\eqn\eightfour{\Delta n/\Delta t = 2 |\beta |^{2} \lbrace 1 \pm n
\rbrace/\Delta t}
where the +(-) is Bose enhancement (Pauli suppression).
The Pauli suppression ensures n(k) $\leq$ 1 for fermions.  Thus to include this
effect we will modify the right hand side of (8.1) by
multiplying by (1$\pm$ 2 f (p,t)). This modified transport eqaution, as we will
show below gives much better agreement with the field theory
calculation.  One can solve the Viasov equation using the method of
characteristics.  From dp/dt = eE and f(p,0)=0 one obtains:
\eqn\eightfive{f(p,t) = \Sigma _{i} \ln [1\pm \exp[ -\pi m^{2}/ |e E(t_{i}) |
]]}
where the t$_{i}$ are determined from
\eqn\eightsix{p+ eA(t) +eA(t_{i}) = 0; t_{i} < t}

The back reaction equation is now
\eqnn\nonumber
$$
\eqalignno{& d^{2}A/dt^{2} = j_{\rm cond} + j_{\rm pol} & \cr
& j_{\rm cond} = 2e \int [dp] p f(p,t) /(p^{2}+m^{2})^{1/2} & \cr
& j_{\rm pol} = 2/E \int [dp] (p^{2}+ m^{2})^{1/2} d^{3}N/dt dx dp & (8.7) \cr}
$$
where
$$
\eqalignno{& dN/dt dx dp = & \cr
& (1\pm 2 f(p,t) ) |e E(t) |\ln [1\pm \exp [ -\pi m^{2}_{\bot}/ |e E(t)
|]]\delta (p) & \cr}
$$
A similar expression holds in boost invariant dynamics as discussed in
\rtwentynine.  The transport approach with the enhancement
(suppression) factor gives reasonable agreement with the direct numerical
solution of the field theory (in lowest order in
1/N) as long as we coarse grain the field theory result in momentum bins.

\newsec{HYDRODYNAMIC CONSIDERATIONS: ENERGY FLOW}

 From a hydrodynamical point of view, flat rapidity distributions seen in
multiparticle production in p-p as well as A-p and
 A-A collisions are a result of the hydrodynamics being in a scaling regime for
the
longitudinal flow.

That is for v=z/t  (no size scale in the longitudinal dimension) the light cone
variables $\tau, \eta$:
\eqn\nineone{z = \tau \sinh \eta; t=\tau \cosh\eta}
become the fluid proper time $\tau=t (1-v^{2})^{1/2}$ and fluid rapidity:
\eqn\ninetwo{\eta= 1/2 \ln [(t-x)/(t+x)] \Rightarrow 1/2 \ln [(1-v)/(1+v)] =
\alpha}

In the rest frame (comoving frame) of a perfect relativistic fluid the stress
tensor has the form:
\eqn\ninethree{T_{\mu\nu} = {\rm diagonal}\; (\epsilon, p, p, p)}
Boosting by the relativistic fluid velocity four vector u$^{\mu}(x,t)$ one has:
\eqn\ninefour{T_{\mu\nu} = (\epsilon + p) u^{\mu}u^{\nu} - pg ^{\mu\nu}}

Letting $ u^{0} = \cosh \alpha ;u^{3} = \sinh \alpha$, we have when $v =zt$
that $\eta = \alpha$, the fluid rapidity. If one has an effective equation of
state $p = p(\epsilon)$ (which happens if both $p$ and $\epsilon$ are functions
of the single variable ($\tau$) as well as for the case of local thermal
equilibrium) then one can formally define temperature and pressure as follows:
\eqn\ninefive{\epsilon + p =Ts;  d\epsilon = T ds; lns = \int d\epsilon /
(\epsilon +p)} Then the equation: $$u^{\mu}\partial^{\nu}T_{\mu\nu}=0$$
becomes:
\eqn\ninesix{\partial^{\nu} (s(\tau) u_{\nu})=0}
Which in 1 + 1 dimensions becomes
\eqn\nineseven{ds/d\tau + s/\tau =0\; {\rm or}\; s\tau = {\rm constant}}

The assumption of hydrodynamical models is that the initial energy
density for the flow can be related to the center of mass energy and a
given volume (say of a Lorentz contracted disk of matter). It is also
assumed that the flow of energy is unaffected by the hadronization
 process and that the fluid rapidity can be identified in the out regime
 with particle rapidity.  Thus  after hadronization the number of pions
 found in a bin of  fluid rapidity  can be obtained from the
energy in a bin of rapidity by dividing by the energy of a single pion having
that rapidty. That is one  assumes
that when the comoving energy density become of the order of one pion/(compton
wave length)\rtwentytwo \rtwentythree we are in the out
regime. This determines a surface defined by
\eqn\eighteight{\epsilon(\tau_{f}) = m_{\pi}/V_{\pi}}
On that surface of constant $\tau$
\eqnn\nonumber
$$
\eqalignno{dN/d\eta &= 1/(m_{\pi}u^{0})dE/d\eta = 1/(m_{\pi} \cosh\alpha) \int
T^{0\mu} d\sigma_{\mu}/d\eta & \cr
d\sigma_{\mu} &= 4\pi a^{2}(dz, -dt) = 4\pi a^{2} \tau_{f}(\cosh\eta,
-\sinh\eta) & \cr
dN/d\eta &= 4\pi a^{2}/m_{\pi} [(\epsilon +p) \cosh \alpha \cosh(\eta
-\alpha)p\cosh \eta]
/\cosh \alpha = 4\pi a^{2} \epsilon(\tau_{f})/m_{\pi} & (9.9)  \cr} $$
 which shows that when $\eta = \alpha$ one gets a flat distribution in fluid
rapidity.  An extra assumption is needed to identify fluid rapidity $\alpha$
with particle rapidity $y=1/2 \ln [(E_{\pi} + p_{\pi}) /(E_{\pi}-p_{\pi})]$,
where $p_{\pi}$ is the longitudinal momentum of the pion.

   What I would like to show next is that in a field theory calculation based
on the Schwinger mechanism if we make the kinematical assumption that the
electric field E is just a function of $\tau$ we obtain a flat rapidity
distribution. We can also prove that the distribution in fluid rapidity is the
same as the distribution in particle rapidity.  We will also obtain
renormalized  expressions for $\epsilon(\tau)$ and p$(\tau)$ (non-equilibrium
dynamical equation of state).

In the pair production problem we have shown that the interpolating phase space
number density is given by the Bogoliubov function (7.7) :
\eqn\nineten{dN/d\eta dk_{\eta} dk_{\bot} dx_{\bot} = |\beta (k_{\eta},
k_{\bot}, \tau)|^{2}}
we are interested in transforming from d$\eta dk_{\eta}$ to dz dy where y is
the particle rapidity $y=1/2 \ln
[(E_{\pi}+k_{z\pi}) / (E_{\pi}-k_{z\pi})]$.

One can show that the transformation from ($\eta, \tau$) to (z,t) is a
canonical one (in the sense of Poisson brackets
 $\lbrace \eta, k_{\eta}\rbrace =\lbrace\tau, \Omega\rbrace =1)$ with canonical
momentum
\eqnn\nonumber
$$
\eqalignno{& k_{\eta}= -Ez +tp = -\tau m_{\bot} \sinh (\eta -y) &\cr
& \Omega =(Et -pz)/\tau) = m_{\bot} \cosh (\eta -y) & (9.11)  \cr}
$$

The phase space is unchanged by this change of variable thus
\eqnn\nonumber
$$
\eqalignno{d^{6}N/(d\eta dk_{\eta} dk_{\bot} dx_{\bot}) &= d^{6}N/dz
dk_{z}dk_{\bot} dx_{\bot} & \cr
&= JdN^{6}/dz dy dk_{\bot} dx_{\bot} & (9.12) \cr}
$$
where J$^{-1} = \partial k_{\eta}/\partial y \partial\eta/\partial z$.  At
fixed $\tau$ one can show that $|J| = dz/dk_{\eta}$
 which leads to desired result, assumed by Landau that along the breakup
surface $\tau$ = constant:
\eqn\ninethirteen{dN/dy = dN/d\eta .}

Schwinger's pair production mechanism leads to an Energy Momentum tensor which
is diagonal in the($\tau,\eta, x_{\bot}$)
coordinate system which is thus a comoving one.  In that system one has:
\eqn\ninefourteen{T^{\mu\nu} = {\rm diagonal}\; \lbrace \epsilon(\tau),
p_{\parallel}(\tau), p_{\bot}(\tau), p_{\bot}(\tau) \rbrace}

We see in a 3 dimensional problem, the field theory in this approximation has
two separate pressures, one in the
longitudinal direction and one in the transverse direction and thus differs
from the thermal equilibruim case.   However,
 for a one-dimensional flow we have that the energy in a bin of fluid rapidity
is just:

\eqn\ninefifteen{dE/d\eta = \int T^{0\mu} d\sigma_{\mu} = A_{\bot} \tau \cosh
\eta \epsilon (\tau)}
which is just the (1 + 1) dimensional hydrodynamical result of (9.9).  This
result does not depend on any assumptions of thermalization.

In the field theory calculation the expectation value of the stress tensor must
be renormalized since the electric field
undergoes charge renormalization.  We can also determine the two pressures and
the energy density as a function of $\tau$.
Explicitly we have in the fermion case.
\eqn\ninesixteen{\epsilon (\tau) = < T_{\tau\tau} > = \tau  \Sigma_{s} \int
[dk] R_{\tau\tau}(k) + E_{R}^{2}/2}
where
\eqnn\nonumber
$$
 \eqalignno{& R_{\tau\tau}(k) =2(p_{\bot}^{2}
+ m^{2}) (g_{0}^{+}| f^{+}|^{2} - g_{0}^{-}| f^{-}|^{2}) - \omega -
(p_{\bot}^{2} + m^{2}) (\pi + e \dot{A})^{2}/( 8
\omega^{5} \tau^{2}) & \cr
& p_{\parallel} (\tau) \tau^{2} = < T_{\eta \eta} > = \tau \Sigma_{s} \int [dk]
\lambda_{s}\pi R_{\eta\eta}(k)
- {1\over 2} E_{R}^{2} \tau^{2} & (9.17) \cr}
$$
where
$$
\eqalignno{R_{\eta\eta} (k) & =2 | f^{+}|^{2} - (2\omega)^{-1}
(\omega+\lambda_{s}\pi)^{-1}&\cr
& - \lambda_{s} e\dot{A} /8\omega^{5}\tau^{2}-\lambda_{s} e\dot{E}/8\omega^{5}
-\lambda_{s}\pi/4\omega^{5}\tau^{2}&\cr
& + 5\pi\lambda_{s} (\pi + e \dot{A})^{2}/(16 \omega^{7}\tau^{2}) & \cr}
$$
and
\eqn\nineeighteen{p_{\bot}(\tau) = < T_{yy} > = <T_{xx} > =(4\tau)^{-1}
\Sigma_{s} \int [dk] \lbrace p_{\bot}^{2}(p_{\bot}^{2} +m^{2})^{-1}
R_{\tau\tau} -2\lambda\pi p_{\bot}^{2} R_{\eta\eta}\rbrace + E_{R}^{2}/2.}

Thus we are able to numerically determine the dynamical equation of state
$p_{i}=p_{i}(\epsilon)$ as a function of $\tau$.

\newsec{DISCUSSION OF NUMERICAL RESULTS}

The physical quantities that we determine numerically are the time evolution of
 E(t), A(t) , and j(t). We will display in the figures the
plasma oscillations and the time scale for field energy to be essentially
transferred into pair production. The other quantities of physical interest we
determine are the spectra of produced particles   dN/dk  and  the dynamical
equation of state.  For comparison we have also solved the phenomenological
transport theory with and without the quantum correction due to statistics
(i.e. Pauli Blocking and Bose Enhancement).  In making plots for the
spatially homogeneous case we use the dimensionless variables \rfive :
$\tilde{E} = (e E/m^{2})\; \tilde{A} = eA/m ; mt = \tau$ When the Electric
field $\tilde E$ is $ > 1$ then it is quite easy for pairs to be produced and
in that regime the final result is independent of the initial data.  We can
see the approach to the tunneling regime by comparing in the regime.  $5 <
\tilde E_{0} < $ 2 the behavior of E($\tau$).  This is shown in fig. 3 for
$\tilde E_{0}$ = .5, 1, 2.  Once $\tilde E_{0} > 2$ the behavior of $\tilde E
(\tau)$ is only weakly dependent on $\tilde E_{0}$.  Once the pairs are
produced one sees that there are plasma oscillations superimposed on which the
electric field degrades.  These figures are from early simulations for scalar
QED in 1 + 1 dimensions \rfive.

In fig. 4 we show $\tilde A (t), \tilde e (t), <j(t)>$ for $\tilde E_{0} = 2$
for scalar QED in 1+1 dimensions \rsix.

In fig. 5 we show $\tilde E$ and $\tilde j$ for scalar QED  in 1+1 dimensions
for $\tilde E_{0}$ = 4. We compare the naive
Vlasov approach (dashed line) and the improved Vlasov approach ( dot- dashed
line) . We notice that including Bose
enhancement corrections is quite important.  We also notice that $\tilde j$ max
= 2 e $\rho$c so that particles continue to
be produced when $\tilde E$  is near a maximum.  In fig. 6 We show the exact
particle spectrum as well as the momentum space
smoothed result which is compared to the Vlasov Equation.  Here $\tilde E_{0}$
=1 and we have scalar QED in 1+1 dimensions.

In fig. 7 we show the results for E and j for $\tilde E_{0}$ = 4 in QED in 1+1
dimension compared to the uncorrected Vlasov
equation. We notice the dismal agreement. In fig. 8 we see the same curves
compared to the improved Vlasov equation. In
fig. 9 we show the exact spectrum of produced pairs for QED in 1+1 dimensions
for $\tilde E_{0}$ = 4.  We notice that n(k)
$\leq$1 which expresses the Pauli Principle. In fig. 10 we compare the binned
version of the field theory result with both
the Naive and Improved transport theory. Next we present recent results for
Scalar QED in 1+1 dimensions using boost
invariant Kinematics.  In fig. 11 we plot E,A and j vs. u=$\log (\tau$) for
E$_{0}(u_{0}$) = 4 in the boost invariant case
where E is a function of the proper time $\tau$ (not to be confused with the
previous $\tau$).  In fig. 12 we compare E(u)
and j(u) with the boost invariant transport theory with and without Bose
enhancement.  Finally we present preliminary
results\ref\r30{Y. Kluger, PhD. Thesis. Tel-Aviv University (June 1992).
(Unpublished)} for scalar QED in 3+1 dimensions.
In
 fig. 13 we show the time evolution for E(t) and j(t) in 3+1 dimensions and
compare with the Boltzmann-Vlasov model with
and without Bose-enhancement.

\newsec{ACKNOWLEDGEMENTS}

The work presented here was done in
collaboration with Emil Mottola, Yuval Kluger, So-Young Pi,
Ben Svetitsky, Judah Eisenberg, Paul Anderson, Barrett
Rogers and M. Samiullah.

\listrefs

\bye